\documentclass[prf,longbibliography,superscriptaddress,twocolumn]{revtex4-2}
\usepackage[T1]{fontenc}
\usepackage[dvipsnames,rgb,dvips]{xcolor}
\usepackage{graphicx}
\usepackage{psfrag}
\usepackage{overpic}
\usepackage{hyperref}
\usepackage{amsmath,amssymb,amsfonts}
\usepackage{mathrsfs}
\usepackage[bbgreekl]{mathbbol}

\newcommand{\ve}[1]{\ensuremath{\mbox{\boldmath$#1$}}}
\newcommand{\ma}[1]{\ensuremath{\mathbb{#1}}}

\newcommand\nn{\nonumber}

\newcommand{\st}{\ensuremath{\mbox{St}}}
\newcommand{\tr}{\ensuremath{\mbox{Tr}}}

\newcommand{\eqnlab}[1]{\label{eq:#1}}
\newcommand{\figlab}[1]{\label{fig:#1}}

\newcommand{\eqnref}[1]{(\ref{eq:#1})}
\newcommand{\figref}[1]{\ref{fig:#1}}

\newcommand{\Eqnref}[1]{Eq.~(\ref{eq:#1})}
\newcommand{\Figref}[1]{Fig.~\ref{fig:#1}}

\newcommand{\Eqsref}[1]{Eqs.~(\ref{eq:#1})}
\newcommand{\Figsref}[1]{Figs.~\ref{fig:#1}}

\newcommand{\algn}[1]{\begin{align} #1 \end{align}}
\newcommand{\pmat}[1]{\begin{pmatrix} #1 \end{pmatrix}}
\newcommand{\sbeqs}[1]{\begin{subequations} #1 \end{subequations}}

\newcommand{\dd}[1]{\tfrac{\text{d}}{\text{d}#1}}

\newcommand{\ee}{\ensuremath{\text{e}}}
\newcommand{\taup}{\ensuremath{\tau_\text{p}}}
\newcommand{\tauk}{\ensuremath{\tau_\text{K}}}

\newcommand{\tth}{\ensuremath{t_\text{th}}}

\newcommand{\co}{\ensuremath{C_\text{O}}}
\newcommand{\cs}{\ensuremath{C_\text{S}}}
\newcommand{\trZ}{\ensuremath{\mathcal{Z}}}
\newcommand{\trAA}{\ensuremath{\mathcal{A}^2}}

\begin{document}
\title{Paths to caustic formation in turbulent aerosols}

\author{Jan Meibohm}
\affiliation{Department of Physics, Gothenburg University, SE-41296 Gothenburg, Sweden}
\author{Vikash Pandey}
\affiliation{TIFR Centre for Interdisciplinary Sciences, Tata Institute of Fundamental Research, Gopanpally, Hyderabad 500046, India}
\author{Akshay Bhatnagar}
\affiliation{NORDITA, Royal Institute of Technology and Stockholm University,
Roslagstullsbacken 23, SE-10691 Stockholm, Sweden}
\author{Kristian Gustavsson}
\affiliation{Department of Physics, Gothenburg University, SE-41296 Gothenburg, Sweden}
\author{Dhrubaditya Mitra}
\affiliation{NORDITA, Royal Institute of Technology and Stockholm University,
Roslagstullsbacken 23, SE-10691 Stockholm, Sweden}
\author{Prasad Perlekar}
\affiliation{TIFR Centre for Interdisciplinary Sciences, Tata Institute of Fundamental Research, Gopanpally, Hyderabad 500046, India}
\author{Bernhard Mehlig}
\affiliation{Department of Physics, Gothenburg University, SE-41296 Gothenburg, Sweden}
\begin{abstract}
The dynamics of small, yet heavy, identical particles in turbulence exhibits singularities, called caustics, that lead to large fluctuations in the spatial particle-number density, and in collision velocities. For large particle, inertia the fluid velocity at the particle position is essentially a white-noise signal and caustic formation is analogous to Kramers escape. Here we show that caustic formation at small particle inertia is different. Caustics tend to form in the vicinity of particle trajectories that experience a specific history of fluid-velocity gradients, characterised by low vorticity and a violent strain exceeding a large threshold. We develop a theory that explains our findings in terms of an optimal path to caustic formation that is approached in the small inertia limit.
\end{abstract}
\maketitle
Ensembles of heavy particles in turbulence, such as water droplets in turbulent clouds \cite{Bod10} or dust grains in the turbulent gas of protoplanetary disks \cite{Joh14,Wil08}, may exhibit large fluctuations of the particle-number density and of their relative velocities \cite{Fal01,Bal01,Bec06,Gus16}. These fluctuations are enhanced by the formation of caustics \cite{Meh04,Wil05,Wil06}, i.e., folds of the particle distribution over configuration space. Caustic formation is an effect of particle inertia, driven by the fluid-velocity gradients, that gives rise to a multivalued particle-velocity field. Due to this multivaluedness, often called the \lq{}sling effect\rq{} \cite{Fal02,Fal07c}, particles may approach each other, and possibly collide, at large relative velocities. Accordingly, caustics have an important impact on the distribution of relative velocities \cite{Gus11b,Gus14c,Per15,Bha18a}, and are a crucial ingredient to theories for collision rates and collision outcomes \cite{Sun97,Vos14,Pum16}. In effect, caustic formation may increase the variance of the particle size distribution in turbulent aerosols because, on the one hand, caustics facilitate particle growth by enhancing collision rates \cite{Gus14c,Bha18a}. Increased collision velocities may, on the other hand, lead to fragmentation, and thus to reduced particle sizes \cite{Wil08}.

Caustics have been extensively studied in direct numerical simulations (DNSs) of particles 
in turbulence \cite{Fal07c} and model flows \cite{Cri92,Mar94,Bec05,Wil05,Duc09,Gus13a,Ree14}.
Recent numerical studies \cite{Per14,Pic19} found that high-velocity collisions tend to occur where the turbulent strain is large, but this cannot be explained in terms of the white-noise models usually used to study caustic formation \cite{Meh04,Wil05,Wil06}. A precise understanding of how caustics form at small particle inertia, including the local flow conditions that lead to their formation, is crucial for the identification of caustics in experiments \cite{Bew13} and for sampling them efficiently in DNS \cite{Fal07c}.

In this Letter, we describe a significant step towards a detailed understanding of how caustics form in turbulence. Using a DNS of two-dimensional turbulence, we show that whether a caustic forms or not depends on the history of the fluid-velocity gradients experienced by closeby particles, not just upon instantaneous correlations between particle positions and flow gradients (preferential concentration \cite{Max87,Squ91,Gus16}). When particle inertia is small, we find a most likely history, i.e., an ``optimal path'' to caustic formation. To determine this path is an optimal-fluctuation problem, similar in nature to localisation due to optimal potential fluctuations in disordered conductors \cite{Zit66}, population extinction due to environmental and population-size fluctuations \cite{Eri13,Kam08}, and shock formation in Burgers turbulence \cite{Gur96,Bec07c}.
\begin{figure}
	\includegraphics[width = 0.7\linewidth]{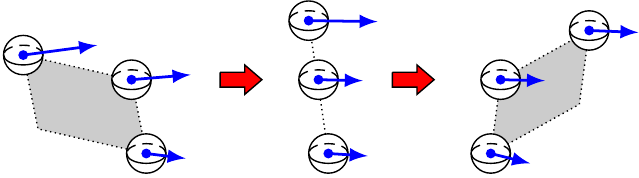}
	\caption{Illustration of caustic formation in two spatial  dimensions. Shown are
	three nearby particles. A caustic is formed when faster particles overtake slower ones.  At a caustic, the area $\mathscr{\hat V}$ of the parallelepiped spanned by separation vectors between the three  particles, shown in grey, vanishes.}\figlab{caust}
\end{figure}
\begin{figure*}
	\includegraphics[width=\linewidth]{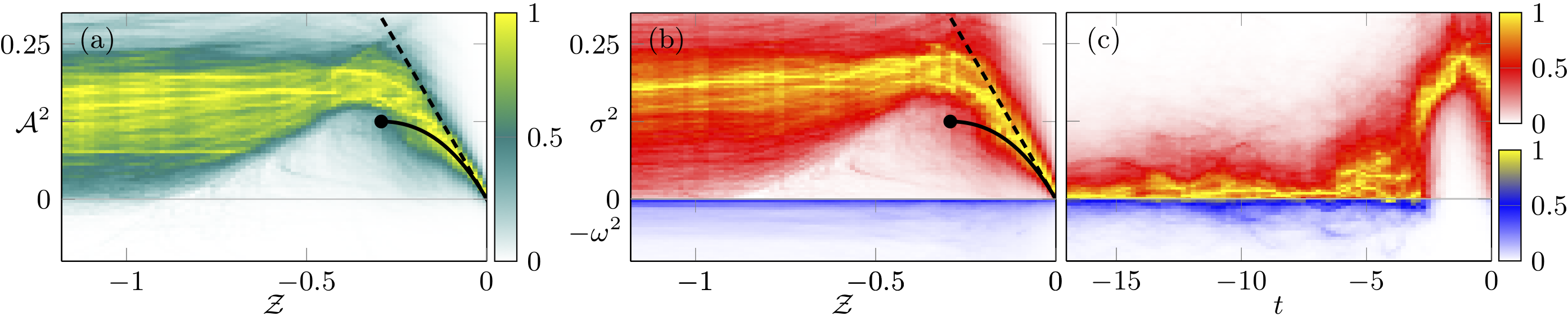}
\caption{Path to caustic formation for $\st = 0.31$. Numerical results using a DNS of two-dimensional turbulence. The path
density is colour coded, normalised to unity along each slice of constant $\trZ$ to improve visibility. (a) Path density in the $\trZ$-$\mathcal{A}^{2}$ plane. The dashed line shows the approximation, $\trZ\sim\trAA$. The solid line shows the evolution of the stable fixed point according to \Eqnref{bifurc}. The dot marks the bifurcation to instability (see main text). (b) Individual contributions from strain ($\trZ$ vs $\sigma^2$, positive axis) and vorticity ($\trZ$ vs $-\omega^2$, negative axis). (c) Path densities for $\sigma^2$ (positive axis) and $-\omega^2$ (negative axis) as a function of time before caustic formation at $t=0$.
}\figlab{smsturb}
\end{figure*}
Based on this observation, we develop a theory that explains how the strain and vorticity change along the optimal path to caustic formation: The fluid strain performs a time-localised, violent fluctuation that exceeds a large threshold, while vorticity remains small. Our results explain qualitatively why DNSs of particles in turbulence show increased collision rates in straining regions  \cite{Per14,Pic19}. Even at finite inertia, the optimal path leaves a clear mark in the data, providing criteria for the identification and the efficient sampling of caustics in experiments and in DNSs.

In a dilute suspension of small, heavy, spherical particles, the dynamics of a single particle is approximately given by Stokes' law \cite{Gus16},
\algn{\eqnlab{stokes}
	\dd{t}\ve x(t) = \ve v(t)\,, \quad  \dd{t}\ve v(t) = \taup^{-1}\{\ve u[\ve x(t),t]-\ve v(t)\}\,.
}
Here, $\ve x$ and $\ve v$ denote particle position and velocity;  $\taup=2a^2 \rho_\text{p}/(9\rho_\text{f}\nu)$ is the particle-relaxation time which depends on the particle size $a$, the kinematic viscosity $\nu$ of the fluid, and the particle and fluid densities, $\rho_\text{p}$ and $\rho_\text{f}$, respectively. The turbulent fluid-velocity field, evaluated at the particle position, is denoted by $\ve u[\ve x(t),t]$.

To describe caustic formation, we consider the parallellepiped spanned by $d+1$ nearby particles in $d$ spatial dimensions. How the spatial  volume $\mathscr{\hat V}(t)$ of this object contracts or expands under the nonlinear dynamics \eqnref{stokes}
is determined by the spatial Jacobian $J_{ij}[\ve x(t_0),t] = \partial x_{i}(t)/\partial x_{j}(t_0)$, namely, $\mathscr{\hat V}(t)= |\det\ma J(t)|$.  Since the dynamics \eqnref{stokes} takes place in $2d$-dimensional phase space, spatial subvolumes $\mathscr{\hat V}$ may collapse in finite time, $\mathscr{\hat V}\to0$, when a caustic forms \cite{Gus16}. Figure~\figref{caust} shows a typical particle configuration that leads to a caustic in two spatial dimensions.
As is well known, caustic formation is closely related to the dynamics of the particle-velocity gradients, which reads, in dimensionless form \cite{Fal02,Gus16},
\algn{\eqnlab{model}
	\st\dd{t}{\ma Z}(t)=-\ma Z(t)-\ma Z(t)^2+ \ma A(t)\,,
}
with initial condition $\ma Z(t_0) = \ma A(t_0)$. Here, the Stokes number $\st = \taup/\tauk$ is a dimensionless measure of particle inertia; $Z_{ij} = \taup\partial v_i(t)/\partial x_j(t)$  and $A_{ij} = \taup\partial u_i(t)/\partial x_j(t)$ are the dimensionless matrices of particle-velocity gradients and fluid-velocity gradients, respectively. In \Eqnref{model}, time is dedimensionalised by the Kolmogorov time, $\tauk^2 = \langle \sum_{i,j=1}^2(\partial u_{i}/\partial x_j)^2 \rangle_s = \taup^2\langle \tr(\ma {A A}^{\sf T})\rangle_s$, where $\langle \cdot \rangle_s$ denotes a steady-state ensemble average. Here and in the following, we use the abbreviations $\ma Z(t) = \ma Z[\ve x(t),t]$ and $\ma A(t) = \ma A[\ve x(t),t]$.  Using \eqnref{model}, one finds  \cite{Gus16} 
\algn{
	\mathscr{\hat V}(t) = \mathscr{\hat V}(t_0)\exp\int^t_{t_0}\!\!{\rm d}s\,\mathcal Z(s)\,,
}
where $\mathcal Z = \tr \ma Z$ is the divergence of the field of particle-velocity gradients. Hence, a necessary condition for caustic formation is that $\mathcal Z$ escapes to negative infinity. 

Apart from the Reynolds number $\text{Re}$ that specifies the turbulence intensity, the particle dynamics is determined by the Stokes number $\st$. For small $\st$, particle detachment is characterised by $\ma A - \ma Z$, which is typically of the order of $\st$, and thus small. 
Caustic formation requires the activation of the nonlinear term in \Eqnref{model} that drives the particle-velocity gradients into a caustic. This, in turn, requires rare and violent fluctuations of the fluid-velocity gradients $\ma A$ of the order of unity.

We determine the dominant events that drive caustic formation by measuring the statistics of paths in the joint space of $\trZ$ and the fluid velocity gradients $\ma A$. In isotropic turbulence, the properties of any statistical quantity must be invariant under rotations. In addition to $\trZ$, we therefore map out the paths of the invariants obtained from the symmetric  ($\ma S$) and antisymmetric  ($\ma O$) parts of the fluid-velocity gradient matrix $\ma A$: $\omega^2 = \tr \ma{OO}^{\sf T}$ and $\sigma^2 = \tr\ma{SS}^{\sf T}$. Figure~\figref{smsturb} shows the paths to caustic formation obtained by numerical simulation, using a DNS of two-dimensional incompressible turbulence. Our simulations are performed in a periodic box forced at the large scales, in the regime of direct cascade of enstrophy. A drag-friction term ensures steady-state turbulence; see the Supplemental Material (SM) \footnote{Supplemental Material available at \ldots for details on mathematical derivations and numerical simulations, as well as for numerical results for the correlation functions in Eq. (9). The Supplemental Material also contains Refs.~\cite{Gra15,Kam07,Mar73,Gra84,Bra89,Ons53}} for more details. The path density in \Figref{smsturb} is colour coded, with the highest densities shown in yellow.

Figure~\figref{smsturb}(a) shows paths to caustic formation in the $\trZ-\trAA$ plane, where $\trAA = \sigma^2-\omega^2 = \tr (\ma A^2)$ is the Okubo-Weiss parameter \cite{Oku71,Wei91} that discerns hyperbolic from elliptic regions in the flow. We see that most paths (yellow regions) that reach a caustic at $\trZ = -\infty$ pass a large fluid-gradient threshold $\trAA\approx 0.2$. The solid and dashed lines are explained in our analysis below.

Figure~\figref{smsturb}(b) shows that typical paths to caustic formation correspond to large strain $\sigma^2$.  Vorticity  $\omega^2$, by contrast, remains small for the majority of paths. Figure~\figref{smsturb}(c) shows the time evolution of $\sigma^2$ and $\omega^2$ prior to caustic formation at $t=0$. We observe that while $\omega^2$ remains small, $\sigma^2$ increases sharply, reaches a large value, and then decreases again. The majority of the large strain, however, persists until the caustic is formed, suggesting that caustics preferentially form in regions of large strain. Appealing to optimal-fluctuation theory, our numerical results point towards an optimal path that underlies caustic formation, characterised by small vorticity and a violent strain. Although the spread in our data is quite large at the value of the Stokes number we used, $\st = 0.31$, the optimal path leaves a strong mark in our data, reflected by the yellow streaks in \Figref{smsturb}.

We explain our observations using an optimal fluctuation approach. The first step is to analyse the fixed-point structure and the bifurcations  of \Eqnref{model}. 
To this end, we expand the equation of motion \eqnref{model} for the $2\times2$ matrix $\ma Z$ in a basis of matrices generated by the identity matrix $\ma I$ and
\algn{\eqnlab{basis}
	\ve e_1 = \pmat{0 & -1 \\ 1 & 0} \,, \quad \ve e_2 &=  \pmat{0 & 1 \\ 1 & 0} \,, \quad \ve e_3 =  \pmat{1 & 0 \\ 0 & -1}\,.
}
This basis is orthogonal with respect to the inner product defined for two matrices $\ma M$ and $\ma N$ by $\langle\ma M,\ma N\rangle=\frac12\tr \!\left[{\ma M} \ma N\right]$, so that $\langle\ve e_i, \ve e_j\rangle = g_{ij} = g^{ij} = \text{diag}(-1,1,1)_{ij}$ and $\ve e^{\sf T}_i = \ve e^{i} = g^{ij} \ve e_j$. Here and in the following, we use the Einstein sum convention. We denote the three-vectors corresponding to $\ma Z$ and $\ma A$ by $z^i(t) \equiv \langle\ve e^i,\ma Z(t)\rangle$ and $A^i(t)  \equiv \langle\ve e^i,\ma A(t)\rangle$, respectively. This formulation in terms of the Lorentzian metric $g_{ij}$ \cite{Nom91} is convenient because it disentangles the strain and vorticity parts of the fluid-velocity gradients $\ma A$. We have $\ma O = \omega \ve e_1$, so that $A^{1} = \omega$ describes the vorticity. The other components, $A^2$ and $A^3$, describe the strain, $\ma S = A^2\ve e_2 + A^3\ve e_3$. Similarity transformations of $\ma Z$, $\ma{\tilde Z} = \ma{ PZP}^{-1}$ leave $\trZ$ invariant, but transform $z^i$ by means of a proper Lorentz transformation, $\tilde z^i = \Lambda^i_j z^j$. The same holds for transformations of $\ma{A}$. The matrix $\mathbb{\Lambda}$ which transforms $A^i$ and $z^i$ has the properties $\mathbb{\Lambda}^{\sf T} \ma g \mathbb{\Lambda} = \ma g$, and $\det \ma{\Lambda} = 1$.

Expanding \Eqnref{model} in the basis \eqnref{basis}, we obtain
\sbeqs{\eqnlab{eomdedim}
\algn{
	\st\dd{t}\trZ &= -\trZ - \frac12\trZ^2 - 2z_iz^i\,,	\\
	\st\dd{t}z^i &= -\left(\trZ+1\right)z^i + A^i\,.
}
}
As the time derivatives on the left-hand side of \Eqnref{eomdedim} are multiplied by $\st\ll1$, we expect the dynamics of $\trZ$ and $z^i$ to take place in the vicinity of its stable fixed points, if they exist. For $A^i = 0$, we find three fixed points, $\trZ=z^i = 0$, $\trZ=-2, z^i = 0$, and $\trZ = -1$, $z_i z^i = 1/4$, whose stability is determined by the eigenvalues of the stability matrix of \eqnref{eomdedim}. The fixed point $\trZ=z^i = 0$ is stable for $A^i = 0$, but a bifurcation occurs at finite $A^i$, $(\trZ,\trAA) = (-1 + 1/\sqrt{2}, 1/8)$, where the fixed point disappears. We conclude that when $\trAA <1/8$, the dynamics \eqnref{eomdedim} takes place in the vicinity of the stable fixed point obtained from the implicit equation
\algn{\eqnlab{bifurc}
	-\trZ(\trZ/2 +1)(\trZ +1)^2 \sim \trAA =  \sigma^2-\omega^2\,.
}
When $\trAA = 2A_i A^i$ exceeds $1/8$, the fixed point ceases to exist, and the nonlinear dynamics \eqnref{eomdedim} drives $\trZ$ to negative infinity, forming a caustic. The evolution \eqnref{bifurc} of the stable fixed point as a function of $\trAA$ and $\sigma^2$ is shown as the solid lines in \Figsref{smsturb}(a) and \figref{smsturb}(b). The fixed points become unstable at the bifurcation point, $(-1 + 1/\sqrt{2}, 1/8)$, (black dots). Hence, for $\trAA<1/8$, particle neighbourhoods are stable and are continuously deformed by the fluid-velocity gradients, according to \Eqnref{bifurc}. For $\trAA>1/8$, however, the neighbourhoods become unstable and collapse after a short time. Expanding \Eqnref{bifurc} for small $\trZ$,  one obtains
the approximation $\trZ \sim -\trAA$ [dashed lines in \Figsref{smsturb}(a) and \figref{smsturb}(b)] used by Maxey \cite{Max87} to explain the preferential concentration of heavy particles in incompressible turbulence \cite{Squ91,Gus16}. This approximation fails to describe caustic formation because it predicts that $\mathcal Z$ remains finite, and thus leads to the incorrect conclusion that particle neighbourhoods are always stable.

Our stability analysis of \Eqnref{eomdedim} explains the qualitative shape of the paths in \Figref{smsturb}(a). However, it misses some of the important results of our DNS. In particular, the stability analysis does not explain why only the strain contributes to caustic formation and vorticity remains small [\Figref{smsturb}(b)], and it has no bearing on the time evolution of the large gradient fluctuations shown in \Figref{smsturb}(c). Finally, we observed in \Figref{smsturb}(a) that the threshold reached by most paths is actually slightly larger than $1/8=0.125$, the value predicted by our stability analysis.

\begin{figure*}
	\includegraphics{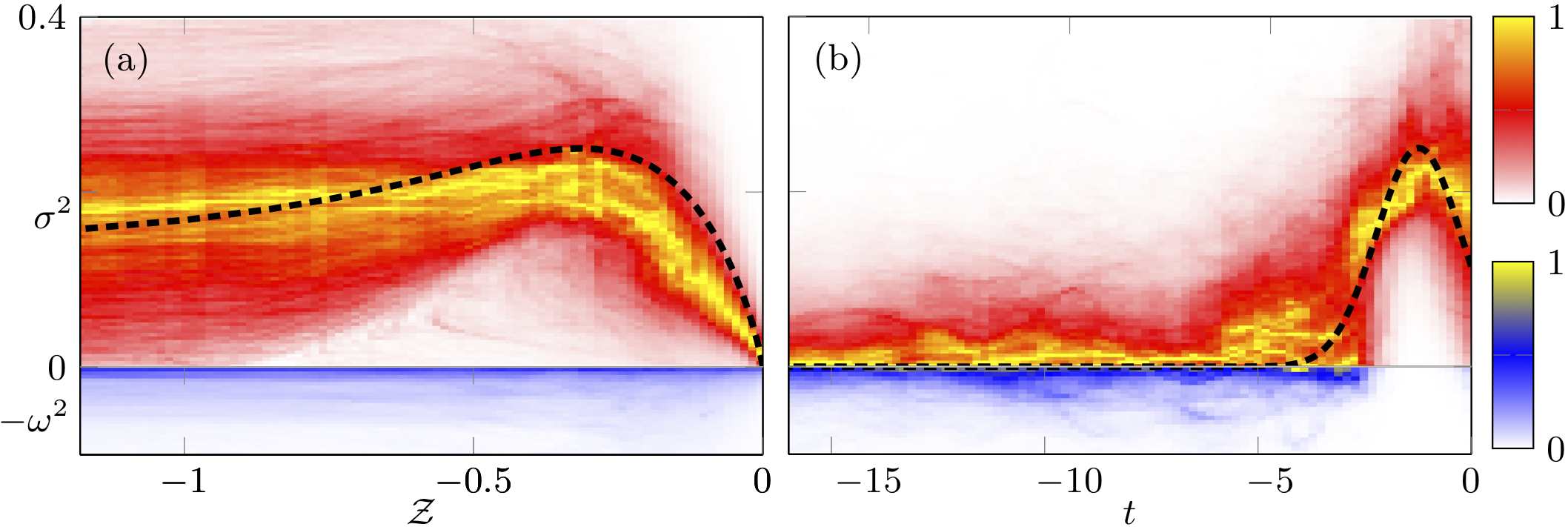}
\caption{
Optimal paths obtained from theory and path density for $\st=0.31$ from DNS. (a) $(\trZ_\text{opt},\sigma^2_\text{opt})$ from theory (dashed line). The colour-coded path density is the same as in \Figref{smsturb}(b). (b) $\sigma^2_\text{opt}(t)$ from theory (dashed line). The colour-coded path density is the same as in \Figref{smsturb}(c).
}\figlab{smstheory}
\end{figure*}

In order to explain these parts of our observations we need to go beyond the stability analysis, and consider how the fluid-velocity gradients reach the large threshold required to render particle neighbourhoods unstable. We do this in the following by computing their optimal fluctuation.

The steady-state correlation functions of $\ma S$ and $\ma O$, evaluated along particle trajectories in isotropic and homogeneous turbulence, have the general form
\sbeqs{\eqnlab{correlation}
\algn{
	\langle S_{ik}(t) S_{jl}(t') \rangle_s &= C^S_{ijkl}\langle \tr \ma{S}(t)\ma{S}^{\sf T}\!(t')\rangle_s \,,	\\
	\langle O_{ik}(t) O_{jl}(t') \rangle_s &= C^O_{ijkl}\langle \tr \ma{O}(t)\ma{O}^{\sf T}\!(t')\rangle_s\,,
}
}
and $\langle S_{ik}(t) O_{jl}(t') \rangle_s =0$.  In two spatial dimensions, the tensors in \Eqsref{correlation} are given by $C^S_{ijkl} = 1/4(\delta_{ij}\delta_{kl} + \delta_{il}\delta_{jk}-\delta_{ik}\delta_{jl})$ and $C^O_{ijkl} = 1/2(\delta_{ij}\delta_{kl} - \delta_{il}\delta_{jk})$. We express \Eqsref{correlation} in terms of the basis \eqnref{basis} to obtain the steady-state correlations of $A^i$,
\algn{\eqnlab{acorrelation}
	\langle A^1(t) A^1(t') \rangle_s &=	\frac12 \langle\tr\ma{O}(t)\ma{O}^{\sf T}(t')\rangle_s\,,\\
	\langle A^2(t) A^2(t') \rangle_s &= \langle A^3(t) A^3(t') \rangle_s =  \frac14\langle\tr\ma{S}(t)\ma{S}^{\sf T}(t')\rangle_s\,.\nn
}
All other correlations of $A^i$ are zero in the steady state. The right-hand sides of \Eqsref{acorrelation} are parametrised as
\sbeqs{\eqnlab{socorrelation}
\algn{
	\langle\tr\ma{S}(t)\ma{S}^{\sf T}(t')\rangle_s = \st^2\cs(\st) f[(t-t')/s]\,,\\
	\langle\tr\ma{O}(t)\ma{O}^{\sf T}(t')\rangle_s = \st^2\co(\st) g[(t-t')/o]\,,
}
}
with the non-dimensional correlation times $s$ and $o$ of $\ma {S}$ and $\ma {O}$, respectively. For tracer particles with $\st=0$ one has $\cs(0) = \co(0) = 1/2$. Inertial particles with $\st>0$ tend to avoid vortical regions due to preferential concentration \cite{Max87,Squ91}, so that $\co(\st)<1/2$. The amplitude $\cs$, on the other hand, 
remains approximately equal to $1/2$ \cite{Per14}. In our two-dimensional numerics, we also find
 $\cs(\st)\approx1/2$ for Stokes numbers between $0.21$ and $0.51$.
The functions $f$ and $g$ in \Eqsref{socorrelation} are normalised to unity, $f(0) = g(0) = 1$. Their time dependencies are well approximated by $f(t) = \exp(-t^2)$ and $g(t) = \exp(-t)$ (see Fig.~1 in the SM \cite{Note1}).

To describe how the fluid-velocity gradients reach the required threshold values, we model  $A^i(t)$ as independent, stationary Gaussian processes with zero mean. For this class of processes, the most probable (optimal) fluctuation $A^i_\text{opt}(t)$ to reach a given threshold can be obtained by optimal-fluctuation methods, as we show in the SM \cite{Note1}. By minimising the action associated with the path probability, we find that the optimal fluctuation of the fluid-velocity gradients is free of vorticity, $\omega^2_\text{opt}=0$, in agreement with our DNSs in \Figsref{smsturb}(b) and (c). This result is intuitive: Vorticity contributes to $\trAA$ with a negative sign, so that any fluctuation of $\trAA = \sigma^2 - \omega^2$ that reaches the large threshold $\trAA_\text{th}$ with finite vorticity requires an even larger strain contribution, to make up for vorticity. The optimal way to reach the threshold value is thus through paths that are vorticity free, whereas the probabilities of paths with finite vorticity are exponentially suppressed. The optimal path for the strain, by contrast, is found to be a time-localised fluctuation, $\sigma^2_\text{opt} = \trAA_\text{th}\ee^{-2(t-t_\text{th})^2/s^2}$, given by a Gaussian function peaked at time $\tth<0$ \cite{Note1}.

Using the optimal gradient fluctuation $(\sigma^2_\text{opt},\omega^2_\text{opt})$, we now obtain the explicit form for the optimal path $(\trZ_\text{opt},\sigma^2_\text{opt})$ as a function of time. For $\st\ll1$, the left-hand sides of \Eqsref{eomdedim} are small most of the time. To evaluate $\trZ_\text{opt}$, we therefore use $A^i_\text{opt}$ as an input into \Eqnref{eomdedim}. Making use of the fact that the vorticity is zero along the optimal path, $\omega_\text{opt}=0$, we find that $\ma A_\text{opt} = A_\text{opt}^i \ve e_i$ and $\ma Z_\text{opt}= 1/2\trZ_\text{opt} +z_\text{opt}^i \ve e_i$ can be brought into diagonal form by a Lorentz transformation \cite{Note1}. The equations for the diagonal entries (eigenvalues) $\lambda^\pm_\text{opt}$ of $\ma Z_\text{opt}$ decouple into two equations,
\algn{\eqnlab{eigenvalues}
	\st\dd{t}\lambda_\text{opt}^\pm = -\lambda_\text{opt}^\pm - (\lambda_\text{opt}^\pm)^2 \pm \frac{\mathcal{A}_\text{th}}{\sqrt{2}}\ee^{-(t-t_\text{th})^2/s^2}\,,
}
with initial conditions $\lambda^{\pm}(t_0) = \pm\mathcal{A}_\text{th}\ee^{-(t_0-t_\text{th})^2/s^2}/\sqrt{2}$. The uncoupled \Eqsref{eigenvalues} are solved numerically, which yields the optimal path $\trZ_\text{opt}$ using $\trZ_\text{opt} = \lambda^+_\text{opt} + \lambda^-_\text{opt}$.

We note that for finite $\st$, the threshold value $\trAA_\text{th}$, determined numerically \cite{Note1}, exceeds the value $1/8$ obtained from the stability analysis, in agreement with our DNS. The reason is that the optimal strain fluctuation $\sigma^2_\text{opt}(t)$ decreases for $t>t_\text{th}$. In order for $\trZ_\text{opt}$ to reach negative infinity, $\sigma^2_\text{opt}(t)$ must exceed $1/8$ for a finite time so that $\trZ_\text{opt}$ can become large and negative. The time for which the threshold must exceed $1/8$ decreases as $\st$ becomes smaller, and we recover $\trAA\to1/8$ in the limit $\st\to0$.

In \Figref{smstheory}, we compare our theory and DNSs. The dashed line in \Figref{smstheory}(a) shows $(\trZ_\text{opt},\sigma^2_\text{opt})$ obtained from \Eqsref{eigenvalues}, with $\st=0.31$ and dimensionless correlation time $s=2.1$ determined numerically. We observe qualitative agreement between the theoretically obtained optimal path and the (yellow) regions of high path density. However, \Eqsref{eigenvalues} slightly overestimate the threshold value $\trAA_\text{th}$. The likely reason is that the Stokes number in our DNS is too large to closely follow our analytical results, valid for $\st\ll 1$.

Figure~\figref{smstheory}(b) shows the time dependence of $\sigma_\text{opt}^2(t)$ obtained from theory with $\st=0.31$ (dashed line) and the corresponding path density from our DNS. We observe that the theory correctly predicts the localised, violent fluctuation of the strain prior to the formation of the caustic, which occurs at $t=0$. A considerable fraction of the large strain required to initialise the caustic persists at
the time of caustic formation, which explains why caustics form preferentially in regions of large strain \cite{Per14,Pic19}.
For other Stokes numbers between $0.24$ and $0.51$, our DNS results lead to the same conclusion (data not shown).

In conclusion, we explained caustic formation in turbulent aerosols at small particle inertia as an optimal-fluctuation problem. In order for caustics to form, the fluid-velocity gradients must follow an optimal path, characterised by small vorticity and a violent strain that exceeds a large threshold. The remnants of the optimal fluctuation at the time of caustic formation result in a strong instantaneous correlation between large strains and caustic events. 

Since caustics give rise to a multivalued particle-velocity field, and thus to high relative particle velocities, our results provide an explanation for the recently observed, instantaneous correlation between particle collisions and intense strain \cite{Per14,Pic19}. 
The characteristic shape of the optimal path to caustic formation will allow one to identify caustics in experiments, and the strong instantaneous correlation of caustics and strain makes it possible to efficiently sample caustics in simulations.

The stability analysis described in this Letter can be generalised to three dimensions, where it reveals that particle neighbourhoods become unstable when the two invariants $Q= -\tr \ma A^2/2$ and $R=-\tr \ma A^3/3$ reach large thresholds in the $Q$-$R$ plane. We therefore speculate that optimal-fluctuation methods also explain caustic formation in three-dimensional turbulence at small Stokes numbers.

\begin{acknowledgments}
K.G. thanks J. Vollmer for discussions regarding the role of the invariants $Q$ and $R$ for caustic formation in turbulence.
J.M., K.G., and B.M. were supported by the grant {\em Bottlenecks for particle growth in turbulent aerosols} from the Knut and Alice Wallenberg Foundation, Grant No. KAW 2014.0048, and in part by VR Grant No. 2017-3865. D.M. acknowledges the support of the Swedish Research Council Grant No. 638-2013-9243 as well as Grant No. 2016-05225. V.P. and P.P. acknowledge support from intramural funds at TIFR Hyderabad from the Department of Atomic Energy (DAE), India and DST (India) Project No. ECR/2018/001135. The simulations were performed using resources provided by TIFR, Hyderabad and the Swedish National Infrastructure for Computing  (SNIC) at the PDC center for high performance computing.
\end{acknowledgments}
\end{document}


\title{Supplemental material {for} \lq{}Paths to caustic formation in turbulent aerosols\rq{}}

\author{{Jan} Meibohm}
\affiliation{Department of Physics, Gothenburg University, SE-41296 Gothenburg, Sweden}
\author{{Vikash} Pandey}
\affiliation{TIFR Centre for Interdisciplinary Sciences, Tata Institute of Fundamental Research, Gopanpally, Hyderabad 500046, India}
\author{{Akshay} Bhatnagar}
\affiliation{NORDITA, Royal Institute of Technology and Stockholm University,
Roslagstullsbacken 23, SE-10691 Stockholm, Sweden}
\author{{Kristian} Gustavsson}
\affiliation{Department of Physics, Gothenburg University, SE-41296 Gothenburg, Sweden}
\author{{Dhrubaditya} Mitra}
\affiliation{NORDITA, Royal Institute of Technology and Stockholm University,
Roslagstullsbacken 23, SE-10691 Stockholm, Sweden}
\author{{Prasad} Perlekar}
\affiliation{TIFR Centre for Interdisciplinary Sciences, Tata Institute of Fundamental Research, Gopanpally, Hyderabad 500046, India}
\author{B. Mehlig}
\affiliation{Department of Physics, Gothenburg University, SE-41296 Gothenburg, Sweden}
\begin{abstract}
In this supplemental material we provide additional details {regarding} the analytical calculations {summarised} in the main {text}, and {for} the numerical simulations {used to obtain the results shown in Figs.~2 and 3} in the main {text}.
\end{abstract}
\maketitle
\section{Instanton calculation for the fluid-velocity gradients}
In this section we explain in detail the instanton calculation outlined in the main {text}. We consider the problem of the gradient process $A^i$ reaching a threshold value given by
$2A_i A^i = 2(-A_1^2+A_2^2+A_3^2) = \mathcal{A}_\text{th}$. In particular, we explain how to derive the time dependence of the optimal paths for the gradient processes $A^i_\text{opt}$, $\omega^2_\text{opt}$ and $\sigma_\text{opt}^2$.

We model the individual components $A^i$ of the fluid-velocity gradient matrix $\ma A$ by independent, stationary Gaussian processes with zero mean and steady-state correlation functions [cf. Eqs. (8) and (9) in the main text],
\algn{\eqnlab{corr}
	\langle A^1(t) A^1(t') \rangle_s &=	\frac12\st^2\co(\st) g[(t-t')/o] \,,\nn\\
	\langle A^2(t) A^2(t') \rangle_s &= \langle A^3(t) A^3(t') \rangle_s =  \frac14\st^2\cs(\st) f[(t-t')/s]\,.
}
Figure~\figref{correlations} shows our numerical data for the correlation functions. Figure~\figref{correlations}(a) shows that $\langle \tr\mathbb{O}(t)\mathbb{O}(0)\rangle_s$ is well approximated by a Gaussian function, $f(t) = \exp(-t^2)$. The correlation function of $\mathbb{O}(t)$, shown in \Figref{correlations}(b), is close to an exponential function, so that we choose $g(t) = \exp(-t)$.
\begin{figure}
	\begin{overpic}{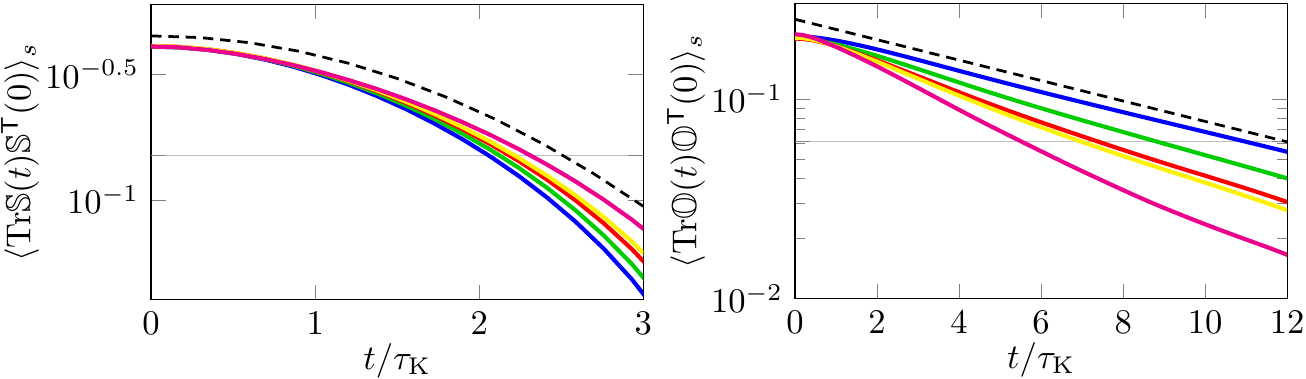}
		\put(12.5,8){(a)}
		\put(62,8){(b)}
	\end{overpic}
	\caption{(a) Steady-state correlation functions for different Stokes numbers, $\text{St} = 0.18$ (blue), $0.24$ (green), $0.31$ (red), $0.34$ (yellow), $0.51$ (magenta), as functions of time, measured along heavy-particle trajectories. (a) Correlation function $\langle \tr [\mathbb{S}(0)\mathbb{S}^{\sf T}(t)]\rangle_s$. The black dashed line shows a Gaussian function for comparison. The light grey line corresponds to $\approx e^{-1}\times \langle \tr [\mathbb{S}(0)\mathbb{S}^{\sf T}(0)]\rangle_s$. (b) Correlation function $\langle \tr [\mathbb{O}(0)\mathbb{O}^{\sf T}(t)]\rangle_s$. The black dashed line shows an exponential function for comparison. The light grey line corresponds to $\approx e^{-1}\times \tr [\mathbb{O}(0)\mathbb{O}^{\sf T}(0)]\rangle_s$.  }\figlab{correlations}
\end{figure}
In \Eqsref{corr}, $o$ and $s$ are the dedimensionalised correlation times defined in the main text. From the gray line in  \Figref{correlations} we estimate $s\approx 2.1$ for all measured Stokes numbers, while $o$ varies between $o\approx 11$ for $\text{St}=0.18$ and $o\approx 6$ for $\text{St}=0.51$.

To obtain the optimal fluctuation to reach a threshold, we consider the transition probability of the gradients $A^i$ from $A^i(t_0) = 0$ at time $t_0$, to a large threshold value $A^i(\tth) = a^i$ at time $\tth$. Because the processes $A^i$ are independent, we compute this transition probability for a single process $A(t)$ with arbitrary correlation function. For the Ornstein-Uhlenbeck process, it is well known \cite{Gra15} how to compute the optimal path to reach a threshold. Here we use a similar method, but generalised to any stationary Gaussian process. The processes we consider need not be Markovian as the Ornstein-Uhlenbeck process, and may not have a representation in terms of a stochastic differential equation.

The transition probability of the zero-mean process $A(t)$ from zero at time $t_0$, to $a$ at time $\tth$ can be written in terms of a path integral
\algn{\eqnlab{pathint}
	P[A(\tth)=a,\tth|A(t_0)=0,t_0] = \mathscr{N}\int\mathscr{D} \left(\frac{z(t)}{2\pi}\right) \exp\left\{i\int_{t_0}^{\tth}\ed t z(t)A(t)\right\}G[z(t)]\,,
}
where $G[z(t)]$ denotes the generating functional of $A(t)$ and $\mathscr{N}$ is a normalisation factor. For Gaussian processes, $G[z(t)]$ is known explicitly \cite{Kam07}
\algn{
	G[z(t)] =& \left\langle \exp\left\{-i\int_{t_0}^{\tth}\ed t z(t)A(t)\right\}\bigg|A(\tth) = a,A(t_0) =0\right\rangle \,,\nn\\
	=& \exp\left\{- \frac12\int_{t_0}^{\tth}\ed t\int_{t_0}^{\tth}\ed t' z(t) c_{t_0}(t,t') z(t')\right\}\,.
}
Here $c_{t_0}$ is the conditional correlation function of $A$, defined as
\algn{
	c_{t_0}(t,t') = \langle A(t)A(t')|A(t_0)=0\rangle\,.
}
The conditioning implies $c_{t_0}(t_0,t)=c_{t_0}(t,t_0)=0$ for all $t\geq t_0$. As $t_0\to-\infty$, $A$ becomes stationary, so that
\algn{\eqnlab{limits}
	\lim_{t_0\to-\infty}c_{t_0}(t,t')=c_s(t-t') = \langle A(t) A(t')\rangle_s\,.
}
Here $c_s(t-t')$ denotes the steady-state correlation function of $A(t)$ which also defines the variance $\eps^2$ of the process, $c_s(0) = \eps^2$. Using the above expressions, we write \Eqnref{pathint} as
\algn{
	P[A(\tth)=a,\tth|A(t_0)=0,t_0] = \mathscr{N}\int\mathscr{D} \left(\frac{z(t)}{2\pi}\right) \ee^{-S_{t_0}[z,A]}\,,
}
with action $S_{t_0}[z,A]$ given by
\algn{\eqnlab{transition}
	S_{t_0}[z,A] = -i\int_{t_0}^{\tth}\!\!\ed t\,	z(t) A(t)
	+\frac12\int_{t_0}^{\tth}\!\!\ed t \int_{t_0}^{\tth}\!\!\ed t' z(t)c_{t_0}(t,t')z(t')\,.
}
We consider weak-inertia case $\st\ll1$ which corresponds to $\eps\ll 1$. In this case, the path integration in \Eqnref{transition} is well-known to concentrate around an \lq{}optimal path\rq{} $(z_\text{opt},A_\text{opt})$ \cite{Mar73,Gra84}. This path is obtained by a variation of the action $S_{t_0}[z,A]$ over $z(t)$ that vanishes at the optimal path:
\algn{\eqnlab{variation}
	0 	= \delta S_{t_0}[z,A]\big|_{A=A_\text{opt},z=z_\text{opt}} = -i\left\{\int_{t_0}^{\tth}\ed t\, \delta z(t)\left[	A_\text{opt}(t) + i\int_{t_0}^{\tth}\!\!\ed t' c_{t_0}(t,t')z_\text{opt}(t')	\right]	\right\}\,.
}
This determines the optimal path $A_\text{opt}$ for $A(t)$:
\algn{\eqnlab{aeqn}
	A_\text{opt}(t) = - i\int_{t_0}^{\tth}\!\!\ed t' c_{t_0}(t,t')z_\text{opt}(t')\,.
}
The transition probability \eqnref{pathint}, and thus the action $S_{t_0}[z,A]$, depends on $A(t)$ only through the final point, $A(\tth)=a$. As a consequence, $z_\text{opt}(t)$ has the general from
\algn{\eqnlab{zeqn}
	z_\text{opt}(t) = i C \delta(t-\tth)\,,\qquad\text{which gives,}\qquad	A_\text{opt}(t) = \frac{C}2c_{t_0}(t,\tth)\,.
}
The constant $C$ is evaluated by using the boundary condition $A_\text{opt}(\tth)=a$. We find
\algn{\eqnlab{aopt}
	\frac{C}{2} = \frac{a}{c_{t_0}(\tth,\tth)}\,, \qquad\text{and thus,}\qquad A_\text{opt}(t) = \frac{a\,c_{t_0}(t,\tth)}{c_{t_0}(\tth,\tth)}\,.
}
Upon substituting \Eqsref{aopt} and \eqnref{zeqn} into \Eqnref{transition}, the action $S_{t_0}[z_\text{opt},A_\text{opt}]$ evaluated along the optimal path $(z_\text{opt},A_\text{opt})$ reads
\algn{
	S_{t_0}[z_\text{opt},A_\text{opt}] = \frac12\frac{a^2}{c_{t_0}(\tth,\tth)}\,.
}
Minimising $S_{t_0}[z_\text{opt},A_\text{opt}]$ over initial times $t_0$ \cite{Gra15} gives $t_0\to-\infty$. This yields for $t\leq \tth$,
\algn{\eqnlab{sza}
	S[z_\text{opt},A_\text{opt}] = \lim_{t_0\to-\infty}S_{t_0}[z_\text{opt},A_\text{opt}] = \frac{a^2}{2\eps^2}\,,\qquad A_\text{opt}(t) = \frac{a \,c_s(t-\tth)}{\eps^2} = \frac{a \,\langle A(\tth)A(t)\rangle_s}{\eps^2}\,.
}
For $t>\tth$, the process is unconstrained and relaxes from $a$ to zero as $t\to\infty$. Unconstrained relaxation does not contribute to the action \cite{Bra89}. Furthermore, for time-reversal invariant processes $A(t)$, the optimal path for the relaxation from the threshold $a$ must be identical to the time-reversed optimal trajectory $A_\text{opt}$ for reaching the threshold. This argument gives $A_\text{opt}$ for all $t$,
\algn{\eqnlab{afin}
	A_\text{opt}(t) = \frac{a \,c_s(|t-\tth|)}{\eps^2}\,.
}
In other words, for a time-reversible Gaussian, stationary process with zero mean, the optimal path for reaching a threshold $a$ is, up to a constant, given by the steady-state correlation function. The action $S$ depends only on the variance $\eps^2$, and is otherwise independent of the details of the process $A(t)$. For the Ornstein-Uhlenbeck process, our results agree with those in Ref.~\cite{Gra15} based on the Onsager-Machlup action \cite{Ons53}.

We now be apply \Eqsref{sza} and \eqnref{afin} to the specific, independent processes $A^i$ with steady-state correlation functions \eqnref{corr} and $f(t) = \exp(-t^2)$ and $g(t) = \exp(-t)$. Comparing \Eqnref{afin} with \eqnref{corr} we have $\eps^2 = \frac12\st^2\co(\st)$, $c_s(|t-\tth|)/\eps^2 = g(|t|)$ for $A^1$ and $\eps^2 =  \frac14\st^2\cs(\st)$, $c_s(|t-\tth|)/\eps^2 = f(|t|)$ for $A^{2,3}$. Hence,
\algn{
	A_\text{opt}^1 =& a^1\exp[-|t-\tth|/o]\,,	\qquad 	A_\text{opt}^{2,3} = a^{2,3}\exp[-(t-\tth)^2/s^2]\,,	\eqnlab{optA}\\
	P(a^i,\tth) =& \lim_{t_0\to-\infty} P\left[A^i(\tth) = a^i,\tth|A^{i}(t_0)=0,t_0\right] \propto \exp\left\{ -\frac{(a^1)^2}{\st^2 \co(\st)} - \frac{2[(a^2)^2 + (a^3)^2]}{\st^2 \cs(\st)}		\right\}	\eqnlab{action}\,.
}
The parameters $a^i$ are obtained by minimising the action $S(a^i)$ in \Eqnref{action},
\algn{\eqnlab{action2}
	S(a^i)= \frac1{\st^2}\left\{\frac{(a^1)^2}{\co(\st)} + \frac{2[(a^2)^2 + (a^3)^2]}{\cs(\st)}\right\}\,,
}
over all $a^1$, $a^2$ and $a^3$ subject to the constraint $2a_i a^i = \trAA_\text{th}$. This gives $a^1=0$, $2[(a^2)^2 + (a^3)^2] = \trAA_\text{th}$ with minimal action
\algn{
	S = \min_{a^i: 2a_i a^i = \trAA_\text{th}}S(a^i)= \trAA_\text{th}/[\st^2\cs(\st)]\,.
}
From this result, we obtain
\algn{
	\omega^2_\text{opt}(t)=0\qquad \text{and}\qquad \sigma^2_\text{opt}(t)= 2[(A^2_\text{opt})^2+(A^3_\text{opt})^2]=\trAA_\text{th}\ee^{-2(t-t_\text{th})^2/s^2}\,,
}
as stated in the main text.
\section{Solution to the eigenvalue equations for $\ma Z$}
We now show how we obtain our solutions for the optimal path for $\trZ  =\tr\ma Z= \lambda^+ + \lambda^-$ for small $\st\ll1$     
 [see main text below Eq.~(10)]. We first explain how to derive Eq.~(10) in the main text. We then briefly discuss our numerical method to determine the threshold value $\mathcal{A}^2_\text{th}$, and to produce Fig.~(3) in the main text. To this end, we write
\sbeqs{\eqnlab{zeqs}
\algn{
	\st\dd{t}\trZ_\text{opt} &= -\trZ_\text{opt} - \frac12\trZ_\text{opt}^2 - 2(z_\text{opt})_iz^i_\text{opt}\,,	\\
	\st\dd{t}z_\text{opt}^1 &= -\left(\trZ_\text{opt}+1\right)z_\text{opt}^1\,,	\\
	\st\dd{t}z_\text{opt}^2 &= -\left(\trZ_\text{opt}+1\right)z_\text{opt}^2 + a^2\ee^{-(t-\tth)^2/s^2}\,,	\\
	\st\dd{t}z_\text{opt}^3 &= -\left(\trZ_\text{opt}+1\right)z_\text{opt}^3 + a^3\ee^{-(t-\tth)^2/s^2}\,.
}
}
At time $t=t_0$ we have $\ma Z(t_0) = \ma A(t_0)$, so that $\trZ_\text{opt}(t_0) = 0$. We conclude that $z^1_\text{opt}(t) = 0$ for all times. Now, we perform a Lorentz transform, which in this case is just a simple rotation, of the remaining components $A_\text{opt}^2$ and $A_\text{opt}^3$ so that $\tilde A_\text{opt}^2 = 0$ and $\tilde A_\text{opt}^3 = \sqrt{(a^2)^2 + (a^3)^2}\ee^{-(t-\tth)^2/s^2} = \mathcal{A}_\text{th}\ee^{-(t-\tth)^2/s^2}/\sqrt{2}$. The leaves the general form of \Eqsref{zeqs} invariant, so that we obtain after the transform
\sbeqs{\eqnlab{zeqs2}
\algn{
	\st\dd{t}\trZ_\text{opt} &= -\trZ_\text{opt} - \frac12\trZ_\text{opt}^2 - 2(\tilde z_\text{opt})_i \tilde z^i_\text{opt}\,,	\\
	\st\dd{t}\tilde z_\text{opt}^2 &= -\left(\trZ_\text{opt}+1\right)\tilde z_\text{opt}^2\,,	\\
	\st\dd{t}\tilde z_\text{opt}^3 &= -\left(\trZ_\text{opt}+1\right)\tilde z_\text{opt}^3 + \frac{\mathcal{A}_\text{th}}{\sqrt{2}}\ee^{-(t-\tth)^2/s^2}\,.
}
}
From $\ma{\tilde Z}(t_0) = \ma{\tilde A}(t_0)$ we again conclude $\tilde z_\text{opt}^2(t) = 0$ for all times. Using the initial decomposition we obtain $\ma{\tilde Z}_\text{opt} = \trZ_\text{opt}/2\ma I + \tilde z^3\ve e_3$, where $\ve e_3$ is diagonal with elements $\ve e_3= \text{diag}(1,-1)$. Hence, we find that after the transformation, $\ma{\tilde Z}_\text{opt}$ is in diagonal form with eigenvalues $\lambda_\text{opt}^+ = \trZ_\text{opt}/2 + \tilde z_\text{opt}^3$ and $\lambda_\text{opt}^-=\trZ_\text{opt}/2 - \tilde z_\text{opt}^3$. Taking time-derivatives of these eigenvalues, and using \Eqsref{zeqs2}, we obtain the uncoupled equations for the eigenvalues $\lambda^+_\text{opt}$ and $\lambda^-_\text{opt}$,
\sbeqs{\eqnlab{eigenvalues}
\algn{
	\st\dd{t}\lambda_\text{opt}^+ = -\lambda_\text{opt}^+ - (\lambda_\text{opt}^+)^2 + \frac{\mathcal{A}_\text{th}}{\sqrt{2}}\ee^{-(t-t_\text{th})^2/s^2}\,,	\\
	\st\dd{t}\lambda_\text{opt}^- = -\lambda_\text{opt}^- - (\lambda_\text{opt}^-)^2 - \frac{\mathcal{A}_\text{th}}{\sqrt{2}}\ee^{-(t-t_\text{th})/s^2}\,,
}
}
presented in Eq.~(10) in the main text. As the trace of a matrix is the sum of its eigenvalues, we have $\trZ_\text{opt} =\tr \ma Z_\text{opt} = \lambda^+_\text{opt} + \lambda^-_\text{opt}$. These equations need to be solved numerically with the initial conditions $\lambda_\text{opt}^+(t_0) = \frac{\mathcal{A}_\text{th}}{\sqrt{2}}\ee^{-(t_0-t_\text{th})^2/s^2}$ and $\lambda_\text{opt}^-(t_0) = -\frac{\mathcal{A}_\text{th}}{\sqrt{2}}\ee^{-(t_0-t_\text{th})^2/s^2}$. 

To determine the threshold value $\mathcal{A}_\text{th}$, we start at $\mathcal{A}_\text{th}=1/8$ and increase $\mathcal{A}_\text{th}$ incrementally until $\lambda^-_\text{opt}$ escapes to negative infinity. By returning to the previous $\mathcal{A}_\text{th}$, and reducing the step size, we obtain an iterative approximation of the smallest value $\mathcal{A}_\text{th}$ for which $\lambda^-_\text{opt}\to-\infty$.

When computing the optimal trajectories shown in Fig.~(3) in the main text, the threshold value $\mathcal{A}_\text{th}$ must be exceeded slightly, in order to enable $\lambda^-$ to escape to negative infinity in finite time. The amount by which the threshold is exceeded is a free parameter in our model. Motivated by the fact that Gaussian fluctuations around the optimal path $A^i_\text{opt}$ are of the order of the square-root of the variance $(\st^2 C_s(\st)/4)^{1/2}$, determined by the action \eqnref{action2}, we choose this parameter equal to $\st/(2\sqrt{2})$. We have checked that our results are insensitive to small changes of this parameter value.%
\section{Direct numerical simulations}
\begin{table}
    \centering
    \begin{tabular}
   {lllllllll}
    \hline\hline\\[-2.25mm]
           $N$ & $f_0$ & $k_\text{f}$ & $\mu$ & $\nu$ &\text{Re}	& $N_\text{p}$ & $\st$ &$\Theta$  \\[1mm]
         \hline\\[-2.25mm]
          1024& $5\times10^{-3}$ & 4 & $10^{-2}$ & $10^{-5}$& 1700	& $10^{5}$ &$0.18-0.51$ & 10 \\[1mm]
          \hline\hline
    \end{tabular}
    \caption{The parameters used in {our} DNS of two-dimensional turbulence. For the numerical evolution we use a time step of $5\times10^{-3}$ units and a total simulation time of $400$ units. The Kolmogorov time evaluates to  $\tauk\approx 2.9$ units.}\tablab{tab}
\end{table}
In this section, we provide some details of the numerical method we used to solve the two-dimensional Navier-Stokes equations and the particle dynamics, resulting in Figs.~2 and 3 in the letter.

In two spatial dimensions, the incompressible Navier-Stokes equations are conveniently expressed in terms of a stream function $\psi(\ve x,t)$ and the fluid vorticity $\omega(\ve x,t) = \ve \nabla \times \ve u(\ve x,t)$. Using the stream function $\psi$, defined by $\ve u = (-\partial_{x_2}\psi,\partial_{x_1}\psi)$, enforces the incompressibility condition $\ve \nabla \cdot \ve u = 0$. In terms of $\psi$ and $\omega = \Delta \psi$, the two-dimensional Navier-Stokes equations read
\algn{\eqnlab{nsvor}
	\partial_t \omega +\partial_{x_1} \psi\partial_{x_2}\omega - \partial_{x_1} \omega\partial_{x_2}\psi  = \nu\nabla^2\omega - \mu \omega + f\,.
}
Here $f$ is a large-scale forcing that generates the turbulence. We take this forcing to be deterministic and time-independent, $f(\ve x, t) = f_0\sin(k_\text{f} x_2)$. The friction term $-\mu \omega$ in \Eqnref{nsvor} ensures a stationary turbulent state. The simulation domain is a square box of side length $L = 2\pi$ with periodic boundary conditions and we discretise the spatial coordinates at $N^2$ points.

We evolve a large number $N_\text{p}$ of particle trajectories by solving simultaneously the Navier-Stokes equation \eqnref{nsvor}, the single-particle dynamics (1), and, along each particle trajectory, the evolution equation (3) for the particle-velocity gradients $\ma Z$. Whenever $\mathcal Z$ exceeds a large negative threshold $\Theta$, we record the formation of a caustic. All simulation parameters are summarised in \Tabref{tab}.